\documentclass[prl,floatfix,twocolumn]{revtex4}
\usepackage{graphicx}
\usepackage{psfrag}

\def\sy{system }
\def\syn{system}
\def\dn{detector}
\def\d{detector }
\def\ds{detectors }
\def\dn{detector}

\def\sc{superconductor }
\def\scn{superconductor}
\def\scg{superconducting }
\def\sa{sapphire }
\def\san{sapphire}

\begin{document}

\title{Fracture Processes Observed with A Cryogenic Detector}

\author{
J. {\AA}str\"om $^6$,
P.C.F.~Di Stefano$^{1,8}$,
F.~Pr\"obst$^1$,
L.~Stodolsky$^{1 *}$,
J.~Timonen$^7$,
C.~Bucci$^4$,
S.~Cooper$^3$,
C.~Cozzini$^1$,
F.~v. Feilitzsch$^2$,
H.~Kraus$^3$,
J.~Marchese$^3$,
O.~Meier$^1$,
U.~Nagel$^{2,9}$,
Y.~Ramachers$^{10}$,
W.~Seidel$^1$,
M.~Sisti$^1$,
S.~Uchaikin$^{1,5}$,
L.~Zerle$^1$
}

\affiliation{
 $^1$ Max-Planck-Institut f\"ur Physik, F\"ohringer Ring
6, D-80805 Munich, Germany;
 $^2$ Technische Universit\"at M\"unchen, Physik
Department, D-85747 Munich, Germany;
$^3$ University of Oxford, Physics Department, Oxford OX1 3RH, UK;
$^4$ Laboratori Nazionali del Gran Sasso, I-67010 Assergi, Italy;
$^5$ Joint Institute for Nuclear
Research, Dubna, 141980, Russia;
$^6$  CSC - IT Center for Science, P.O.Box 405, FIN-02101 Esbo, 
Finland;
$^7$ Department of Physics, P.O. Box 35 (YFL), FIN-40014 University
of Jyv\"askyl\"a, Finland;
$^8$  Institut de Physique Nucl\'eaire de Lyon, Universit\'e Claude
Bernard Lyon
I, 4 rue Enrico Fermi, 69622 Villeurbanne Cedex, France;
$^9$  Institute of Chemical Physics and
Biophysics, EE-0026 Tallinn, Estonia;
$^{10}$University of Warwick, Dept. of Physics, Coventry CV4 7AL,
UK,
$^*$ Corresponding author, {\it email address:}  les@mppmu.mpg.de.}

\begin{abstract}
 In the early stages of  running of the CRESST  dark matter
search
using sapphire  \ds at very low temperature,
an unexpectedly
high rate of signal pulses appeared. Their origin was finally
traced to 
 fracture events  in the sapphire due to the very tight clamping of
the
detectors. 
During extensive runs the 
energy and time of each event was recorded, providing  large
data
sets for such phenomena.  We believe this is the first time the
energy release in  fracture has been directly and accurately
measured on a
microscopic
event-by-event basis. The energy threshold  corresponds to 
the breaking of only a few hundred covalent bonds, a sensitivity
some orders of magnitude greater than that of previous technique. 

We report some features of the data, including
   energy distributions, waiting time distributions,
autocorrelations and the Hurst
exponent. The
energy distribution appear to follow a power law,  $dN/dE\propto
E^{-
\beta}$, similar to the 
 power law for earthquake magnitudes, and after
appropriate translation, with a similar exponent. In the time
domain,    the waiting
time $w$ or gap distribution between events   has a power law
behavior at
small $w$ and an exponential fall-off at large $w,$ and can be
fit 
  $\propto w^{-\alpha}e^{-w/w_0}$. The
autocorrelation
 function shows  time
correlations
lasting  for substantial parts of an hour. An asymmetry is found
around large events, with higher count rates after, as opposed to
before, the  large event .
\end{abstract}

\maketitle
\section{Introduction}
 In the Spring of 1999 preliminary runs of the CRESST
dark matter search~\cite{cresst}  were carried out at  the Gran
Sasso Laboratory
(LNGS), a deep underground laboratory for low background physics
located in the Apennines. In these
first runs of CRESST a
phenomenon was observed which we believe may be of interest for the
study of crack and fracture formation in brittle materials.
CRESST  is a cryogenic  detector, working
in the vicinity of 10 milli-Kelvin~\cite{leo}. In addition to being
deep
underground  for shielding against cosmic rays, it is carefully
designed to minimize effects of
radioactive background. The  \d elements were
large (262 gram) high quality single  crystals of sapphire, with a
strip of \sc (W) evaporated on one
surface to serve as a sensitive thermometer.
This system, as  shown by tests with gamma ray  sources, detects
single events in the \sa with energies in the range  from about 1
keV to several
hundred keV with good energy
resolution ( 0.5 keV) and good time resolution (40 or 100 $\mu$s
for the onset of a pulse).

 In order to reach these low temperatures it is important to
eliminate the effects of any vibrations (``microphonics'') that
might deliver energy to the crystal. Thus in addition to special
suspensions to isolate the apparatus, 
the crystals are held very tightly in their holders to prevent any
even microscopic
frictional effects.  In the data to be discussed
 here this  was effected by  small  \sa balls held against
the \sa crystal by a plastic clamp.  The plastic of
the clamp, delrin,  is known to
contract substantially at low temperature, thus providing
additional
``tight holding''.   An unanticipated result of
the small contact area of the hard \sa balls and the great force of
the
clamp turned out
to be a cracking or fracturing of the \san. This was observed
as follows.

 When the \sy was first
brought into operation, an unexpectedly high
rate of signal pulses was observed.  Initial fears that this might
be due to an unexpected radioactive contamination were relieved by
the observation that even an unknown radioactive contamination must
be Poisson distributed in time, while the unexpected pulses 
appeared rather  to come in ``bursts" or ``avalanches''.
Examination
of the time
 distributions showed that they were indeed non-Poissonian.

{\it  Pulse formation and fractures:} 
 The pulses themselves resembled those seen from good
particle events. However, this is a rather unspecific criterion,
due to the operating characteristics of the \dn.  There  are
essentially three steps in the production of a signal pulse 1) A
relatively localized
energy release  within a short time,  2) A rapid
degradation
of this energy into a uniform ``hot'' ($\sim 10\, ^oK$) gas of
phonons produced  through  phonon-phonon
interaction and decay, as well as interaction with the crystal
surface, 3) Absorption of the phonons in the
thermometer strip. This 
leads to a heating with an increase of electrical
resistance
for the \scn, which is finally read out by SQUID electronics.   
The
resulting pulse shape is  well described by a model employing the
various
thermal and electrical parameters of 
the \sy\cite{model}.
 As may be seen from this brief description, the pulse shape is
essentially determined by the thermal responses of the system and
not by the initiating event, as long as it is ``fast''.
Hence any  release of a given energy in the crystal
 in a short time ($\mu$ seconds)  leads to the same 
pulse shape and so examination of  the pulses does not lead to an
identification  of their origin.
 An extensive search for the origin of the pulses was finally
successful
when it was noticed that there appeared
to be markings or scratches on the crystal at the contact points
with the \sa balls. When  the  \sa balls were replaced by
 plastic    stubs, which are   evidently much softer,   the
event rate  immediately dropped from some thousands  per hour to
 the    expected few per hour.  

 These observations strongly suggest that the  pulses
 were due to some kind of cracking or
micro-fracturing phenomena in the \sa crystal and/or its support
balls.
Indeed, examination under a microscope revealed a small crater with
radiating irregular fissures extending sideways and down into the
crystal. Damage to the sapphire balls was also observed.
Since the reduction in rate after the exchange of  the \sa balls  
 was so large, we believe the
data taken with the \sa balls are essentially all  fracture
events. 
If we accept this  crack or fracture hypothesis, our data then
represent a  large 
sample of well measured fracture events, under low background
conditions, and with good time and energy determination. 

{\it Calibration runs:} In order to calibrate the energy scale
regular
calibration runs were carried out. In these runs the system is left
undisturbed  and a radioactive  source supplying 120 keV photons
(which can
penetrate to the detectors) is inserted in an external plug in the
shielding. These
photon-induced events can be selected by using the resulting 120
keV
peak in the data. Since a radioactive source produces statistically
independent events, that is Poisson statistics,
 these events  provide a useful  comparison when 
studying   statistical properties of the data.

\section{Energy distributions}
 We believe this is the first time that the energy release in
microfracture  has been  accurately measured on a microscopic
event-by-event basis.

  It is to be emphasized that the cryogenic method
 provides an {\it absolute} measurement of the {\it total}
energy release in the fracture. This is to be contrasted with
the study of acoustic
emission in materials or seismic measurements of earthquakes. There
the
energy determination is necessarily indirect since there are
various
assumptions and uncertainties concerning production, propagation,
and
detection   involved in translating the observed signals
into the true energy of the event. On the other hand the cryogenic
method,  essentially
calorimetric in character,  is a direct measurement of the full
energy. The
energy scale is  fixed by the calibration with known sources and 
the
resulting accuracy of the CRESST energy determination is on the
order of a few  percent~\cite{cresst}.

In addition to the directness of the  energy measurement    an 
important feature of the cryogenic  method is its great
sensitivity.
The closest previous technique appears to be the study of
acoustic
emission in materials ~\cite{R1},\cite{rogers}.  There the
smallest emitting
region  considered is on the order of a  square micron
~\cite{rogers1}. This will correspond to the breaking of  
 $\sim 10^{7}$ bonds in the crystal. On the other hand,
our energy threshold is typically some keV (Fig. 1). This
corresponds  to
the breaking of only a  few hundred or  thousand  bonds. Thus  the
cryogenic method appears to be many orders of magnitude more
sensitive than  previous technique.
 Small cryogenic devices can even be  sensitive to energies in the
eV range~\cite{leo} and it is possible that studies of this type 
involving stress release of only a few atoms  are
feasible~\cite{me1}.

 In Fig.~1 we show the differential distribution $dN/dE$ for the
number of events $N$  
per unit energy,  for four data sets with two detectors from  Run9.
The  straight line is the result of a power law fit 
\begin{equation}\label{e}
dN/dE\propto E^{-\beta}
\end{equation}
to the lowest
curve, which yields $\beta \approx 1.9$. Similar results  are found
from
fits to other data sets. From
  a total of seven sets  examined
(from Runs 9, 10 and 11)  $\beta$ 
 ranged  between $ 1.7$ and $2.0$. An interesting point is 
that the rates do not appear to differ greatly from one data set to
another,
despite the fact that different crystals and mountings are often
involved. At  21 keV for example, the rates   over the
various data sets vary between 4 and 11 pulses/keV-hr.

\begin{figure}[h]
{{\includegraphics[width=\hsize]
{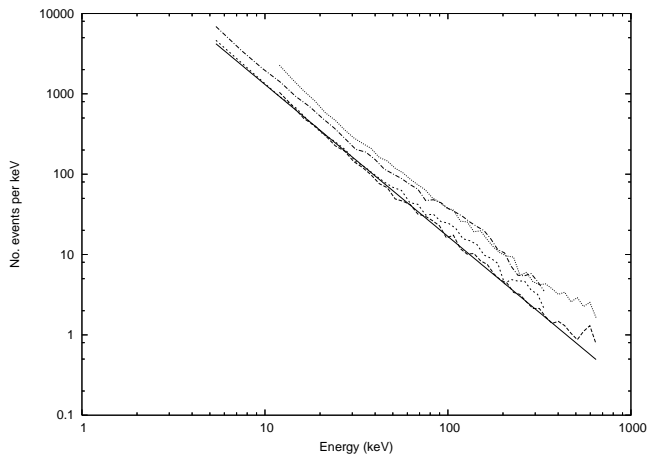}}}
\caption{ Energy spectra from four data sets of Run9, with 53 hr
for the upper pair of curves and 28hr for the lower.  The straight
line shows  a  fit to the lowest curve $ \propto
E^{-\beta},$ yielding $\beta \approx 1.9$. }
\end{figure}
A power law of this type, called
the Gutenberg-Richter law~\cite{gr}, is well known for the
``magnitudes''
of earthquakes. Unfortunately  the ``magnitude'' is a seismic
amplitude and not a direct measurement of the energy of an
earthquake. Thus a simple comparison is not possible.
 However if one takes the  prescription that the seismic
amplitude to approximately the 3/2
power~\cite{gr},\cite{bak} represents the energy,
and uses the power $\approx 1.0$ found for the {\it integral}
distribution of earthquake magnitudes~\cite{bak},  
it corresponds  to $\beta \approx 1+{2\over 3}\approx 1.7$, not
 far from our  $\beta \approx 1.7-2.0$. Of course,
 the six orders of magnitude range available for 
 seismic data is much greater than the one or two orders of
magnitude available here. 

  It should also be noted that such power
law, that is scale free, distributions appear in many phenomena,
often related to an underlying fractal process~\cite{west}. In 
the acoustic emission recordings of
microfracture events in brittle 
 materials for example, such a distribution typically appears, with
a somewhat lower exponent,
 $\beta\approx 1.5$~\cite{R1}.

\section{ Time Series  }
{\it Waiting Time Distributions:} A useful quantity in the study of
intermittent data such as the present is
the ``waiting time'' $w$. To each event $i$ we
assign
 $w_i$, the time interval till the next event, and study the
distribution 
of these intervals. Fig.~2  shows the waiting time distribution for
detector
2 in a
28 hr data set of Run 9.  The distribution has power law behavior
at small $w$ and an exponential fall off at large $w$, and an
accurate   fit is obtained  with  $ dN/dw\propto
w^{-\alpha}e^{-w/w_0}$, with $\alpha=0.33$.
 Similar
results are
found for other data sets with  $\alpha$ in the range 0.25-0.5.
 The parameter $w_0$
determines the location of the crossover from power law to
exponential and is essentially the inverse rate or average waiting
time, with ${\bar w}=(1-\alpha)w_0$.  Qualitatively similar
results, with 
   $\alpha$ near to or somewhat less than one, have 
been reported   for earthquakes in California~\cite{R2}.

\begin{figure}[h]\label{ww}
{{\includegraphics[width=\hsize]
{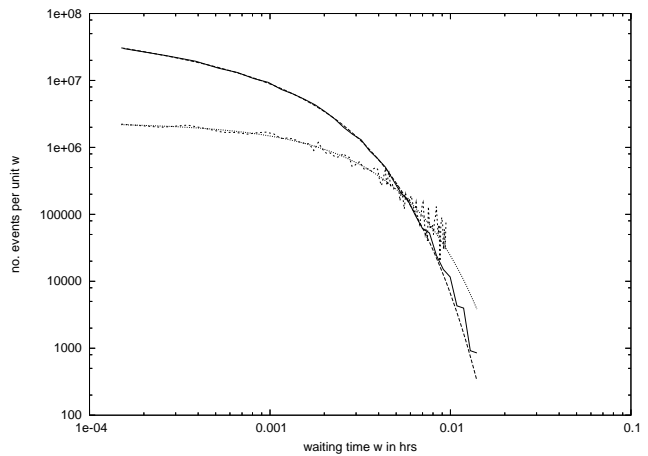}}}
\caption{Waiting time distributions. Upper curve: fractures,
  fit to $\propto w^{-\alpha}e^{-w/w_0}$. Lower curve:
 photon-induced events from  a calibration run,  fit to
$\propto e^{-w/w_0}$.}
\end{figure}
For the simple case of Poisson statistics,  one expects a waiting
 time distribution
 $\propto e^{-w/w_0}$, where $1/w_0$ is the
average count rate. The lower curve of Fig.~2 shows the waiting 
time 
distribution for the photon-induced events of a calibration run,
with a fit
to $\propto e^{-w/w_0}$. 
As expected there is a good fit, and with $1/w_0$  in agreement
with the
 event rate.

An interesting point concerns the behavior of $w_0$ for fracture
 events as the 
 the energy threshold for the sample is raised. It
appears
that the form
$w^{-\alpha}e^{-w/w_0}$ is preserved, with $\alpha$ varying little.
Since
 the count rate is reduced
however,
the value of $w_0$ increases and so the crossover between  power
law
and exponential behavior moves to larger $w$.
 Indeed, taking a given data set (Run9-d2,$100 \mu s$), repeatedly
raising 
the energy  threshold  and fitting for 
$w_0$, we find a linear relation between the
inverse
 count rate,
that is $\bar w$, and the fitted $w_0$.
   The slope
and the relation ${\bar w}=(1-\alpha)w_0$ then gives a global
determination  $\alpha\approx
0.26$.

 The power law behavior for the waiting times at small $w$, as well
as the power law for the
energy distribution in the previous section, is suggestive of an
underlying scale-free processes without any intrinsic dimensional
parameter, as is common in fractal processes~\cite{west}. However
this cannot be entirely true here since $w_0$
is a
time and has 
dimensions.  Since  $e^{-w/w_0}$
corresponds in fact to a  Poisson distribution,
 this may
suggest an interpretation in terms of some basic scale free
processes
where several such processes are occurring
independently and simultaneously and so are overlapping in the
data.
This arises trivially if the signals originate from more than one
of the
 support points of the crystal, of which there were several; but
one can also
 imagine independent crack systems beneath one support point.

The increase of $w_0$ as the count rate goes down
suggests that the
limit of zero count rate is a kind of critical point:  the 
waiting
time becomes infinite as the
distribution becomes non-integrable and  completely scale free,
while  $1/w_0$ appears as a diverging correlation length.
Understanding $w_0$ is an interesting point for further study.

{\it  Correlations in Time:}
 We expect  the
existence of correlations in time, corresponding to the
``bursts" or ``avalanches''.   We use the event
rate
$R_t$    of a
calibration run to
construct  the autocorrelation function 
\begin{equation}\label{au}
C(t- t')=\overline{(R_t-{\bar R})(R_{t'}-{\bar R})}\; 
\end{equation}
and 
compare $C$ for photon-induced events  and fractures in Fig.~3.
While for photons we have $C=0$ as expected,
    for the microfractures  there are
correlations lasting for substantial fractions of an hour. 
These  long-term correlations are found for the
 fracture events of all data
sets. The physical origin of the correlations may be in stress
relaxation
   phenomena where a slow ``diffusion" of strain~\cite{R3} can
trigger new
   microfractures when  meeting other weak spots
   in the crystal.

\begin{figure}[t]
{{\includegraphics[width=\hsize]
{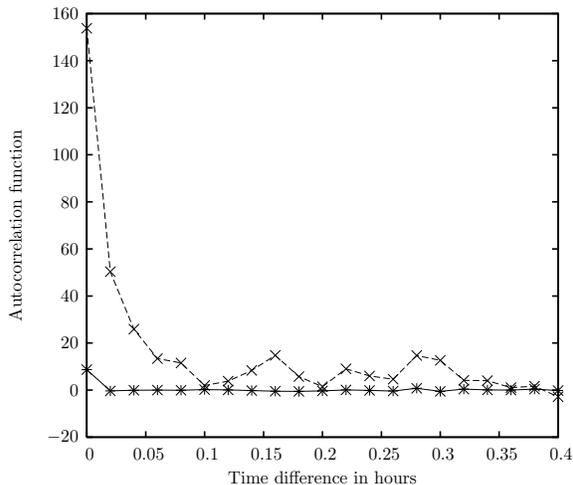}}}
\caption{Autocorrelation function $C$ for the event rate 
 from  calibration data  of Run10. The lower
curve is
for photon-induced events (events in the 120 keV peak), and the
upper curve
is for fractures plus some admixture of compton scatters (events
below the peak). For the photons the data
is
consistent with $C=0$ for $(t-t')\neq 0$ as expected for Poisson
statistics, 
with $C(0)=Variance= {\bar R} $.}  
\end{figure}

{\it Hurst exponent:}  The autocorrelations as in Fig.~3 can be
approximately fit to power laws $\propto (t-t')^{-p}$. As noted
above, this is  suggestive of the
  scale free, self-similar behavior associated with fractal
statistics.
 A way of
characterizing such behavior is in terms of what is called the
Hurst exponent
H; and  we can check the plausibility of such a description by
comparing the
consistency of  H found in
different ways.  Table I shows H found in three ways for various
data sets. First the autocorrelation exponent
$p$ is fitted to find $H=1-p/2$. The next column
shows $H$ determined 
by  the ``growth of the standard deviation",  a
characterization  of the fluctuations in the event rate $\sim t^H,$
  where $t^{0.5}$ would be 
 the classical gaussian or  random walk  behavior with
 finite range correlations. Finally,  the last column  gives $H$ 
found
from the 
``Shannon Entropy'', related to the probability of the number of
events over a time
interval $t$~\cite{scafetta}.  Although the
fits were not all excellent and there is
considerable fluctuation in the results, the overall rough
consistency of  the three determinations supports the
picture of a scale free,  self-similar process. We do not
necessarily expect the same $H$ for different data sets since these
involve different energy thresholds and sensitivities.

\begin{table}
\caption{Exponent H found by different methods.
 d1 and d2  refer to the two \ds in operation,
and 40,
100 $\mu s$
 to different digitization windows used in 
data taking in Run9.}

\begin{tabular}{|l|l|l|l|} 
\hline
Data Set&Autocorr.& Stnd. Dvtn. &Sh. Entropy\\
\hline
\hline
Run9 d1 100$\mu s$&0.77&0.70&0.69\\
\hline
Run9 d2  100$\mu s$&0.80&0.80&0.80\\
\hline
Run9 d1 40$\mu s$&0.73&0.70&0.67\\
\hline
Run9 d2 40$\mu s$&0.69&0.70&0.65\\
\hline
Run10 d2&0.59&0.63&0.59\\
\hline
Run11 d1&0.60&0.64&0.53\\
\hline
Run11 d2&0.69&0.66&0.62\\
\hline
\end{tabular}
\end{table}

\section{ Clusters}
 A frequently used concept in the earthquake
literature is the ``Omori Cluster": a ``big shock'' followed by
``aftershocks''. As Fig.~1 shows, and as is also the case
for earthquakes, there
is no separate class of  high energy events---no distinctive ``big
shocks''. Naturally, as
 should  be expected from the ``avalanches'' or correlations,
given any event, there is a general increase in  rate at nearby
times.
 Although this increase is quite substantial, (a factor
four
with one second bins, see
 Fig.~4) this simply reflects the ``bursts" or ``avalanches" and is
not   specific to ``big events''.

  More specific to   ``big events'', however,  we  find a  time
asymmetry with respect to ``before'' and ``after''. That is, there
is on average more activity  after,
as
opposed to before,  ``big events''.
  Fig.~4 shows the  count rate from a data set of Run9,   for times
 close to ``big
events'', with   ``after'' (upper histogram) and ``before" (lower
histogram)
  plotted separately.
``Big'' was defined as a  pulse with $E>300keV$. The  bin
 size is $2.0\times 10^{-4}$hr$=0.72$s.
One notes a significantly higher rate in the first bin ``after''
relative to that in the first bin ``before''. There is a reasonable
fit to a power law for the decline in the rate toward the average
value, and  a  significantly steeper power ``after'' relative to
that
``before'' is found.  Similar
results are  obtained for other data sets. 
  
\begin{figure}[h]\label{elv}
{{\includegraphics[width=\hsize]
{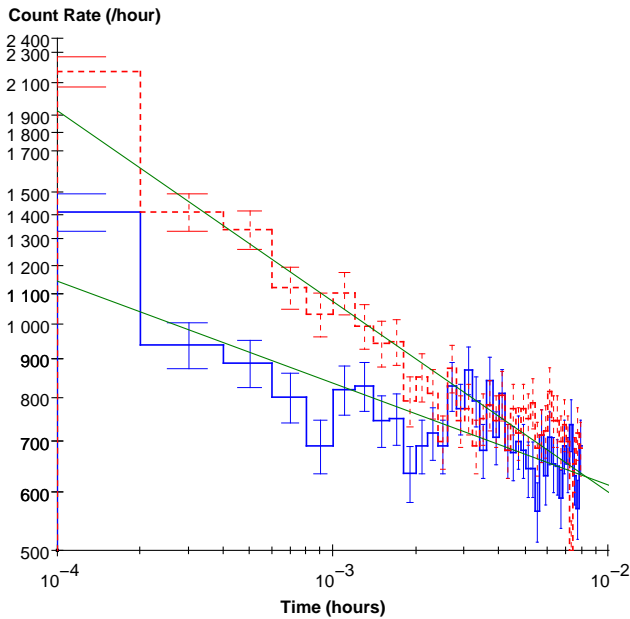}}}
\caption{ Count rates in the vicinity of ``big events'', showing
a time asymmetry ``before'' and ``after'' the ``big events". From
a
53
hr data set of Run9, plotted in 0.72 s bins. The upper (dotted,red)
histogram is for times following the ``big event'' and the lower
(solid,blue) histogram for times preceding the ``big event''. The
straight lines are power law fits, yielding a power $0.13 \pm 0.01$
``before'' and a power $0.25 \pm 0.01$ ``after''. There were 
1082 ``big events'',  defined as a single pulse with $E>300keV$. 
 The average rate in the run  was 526/hr. }
\end{figure}

An asymmetry of this type appears to exist in seismic data and
certain models~\cite{Helm}
and seems to indicate that the
 ``big events" tend to occur early in the ``bursts".

\section{ Crack Propagation and Material Properties}
 Our material
is a single crystal of high purity\cite{pure}.   In 
  crack propagation models the growing stress enhancement
at the crack tip implies that a ``hard spot'' is necessary to limit
the propagation 
of a crack; thus when a
homogeneous stress is applied to
  a defect-free material there is nothing to stop a propagating
crack.
Presumably the microfractures here were limited by  
the random,
non-homogeneous stress and defect
field which quickly arises as  fractures  form in the pure
material. This may
have been assisted by the
damage to
the
small sapphire
balls,  leading to an
irregular 
application of the stress. Although we speak of ``cracks'', it
should be kept in mind that from our simple observation of  pulses
we cannot infer the exact nature of the microfracture. Finally,
with
respect to materials it should be noted that our system is of
course quite opposite to those in the geological context, where one
 has highly heterogeneous systems, while here we have a very pure
material.

\section{ Development of the Technology}

 It is interesting to contemplate the extension of this method
in the study of fracture phenomena.
The \scg thermometer, and perhaps other cryo-sensors~\cite{leo},
 can be applied to many materials.
The very low
temperature and large crystals of the dark matter search would
 not 
always be needed, and indeed it might be possible to follow  the
crack development in time with a 
 smaller and  thus faster \syn.  However, low background
conditions  may still be necessary  to
avoid contamination of the
data by non-fracture events. 
 In the present data  the crystal was contacted by several
of the small sapphire balls, and we are unable to determine where
an event originates. Such effects lead to a dilution of
correlations, which thus
may be intrinsically much stronger than appear here. In an 
apparatus
especially designed
for such studies
one could arrange to have only one ``hard'' contact and with  a
known
force. Finally, since the energy range available is relatively
small compared to  that for earthquakes it would be useful to
consider techniques for
increasing  the  dynamic range.


\begin{thebibliography}{00}
\bibitem{pure} The crystals were high quality  single crystals,
optically defect free. One was supplied by the Hemex company, the
other by BEC Breznikar.
\end{thebibliography}
\end{document}